\documentstyle[aps,psfig,pra,multicol]{revtex}
\begin{document}

\title{Transparency near a photonic band edge}
\author{E. Paspalakis, N.J. Kylstra and P.L. Knight}

\address{Optics Section, Blackett Laboratory, 
Imperial College, London SW7 2BZ, United Kingdom}

\date{\today}
\maketitle
\begin{abstract}
We study the absorption and dispersion
properties of a ${\bf \Lambda}$-type
atom which decays spontaneously near the edge of a
photonic band gap (PBG). Using an isotropic PBG model, we show that
the atom can become transparent to a probe laser field,
even when other dissipative channels are present. 
This transparency originates from the square root singularity
of the density of modes of the PBG material at threshold.
\end{abstract}

\begin{multicols}{2}
The study of quantum and nonlinear optical phenomena 
in atoms (impurities) embedded in 
photonic band gap (PBG) materials has attracted
much attention recently. Many interesting effects have been predicted 
when radiative transitions of the atoms are near-resonant
with the edge of a PBG. As examples we mention the 
localization of light and the formation of 
`photon-atom bound states' \cite{John87a,John90a,John95a}, 
suppression and even complete cancellation of spontaneous emission \cite{Yablonovitch87a,Quang94a,Kurizki94a,Bay97a}, population trapping in 
two-atom systems \cite{Bay97a}, phase dependent behaviour of the population dynamics \cite{Quang97a}, enhancement of spontaneous emission interference \cite{Zhu97a} and other phenomena \cite{Bay97b,Vats98a}. 
In addition there is also current interest with regard to
quantum non-demolition measurements in modified reservoirs, such as the PBG \cite{Kofman96a,Lewenstein99a}. We note that there is 
a formal similarity between the models used in the above 
studies and those of near threshold photoionization and photodetachment \cite{Eberly82a,Bhatt88a}.

In this Rapid Communication we study the probe absorption spectrum of a ${\bf \Lambda}$-type
system, similar to the one used in previous studies \cite{Quang97a,Bay97b}, 
with one of the atomic transitions decaying spontaneously near
the edge of a PBG. We show that the atom becomes transparent
to a probe laser field which couples to the second atomic transition. 
This transparency occurs even in the presence of 
the background decay of the upper atomic level.
This effect is closely related to the phenomenon of
electromagnetically induced transparency (EIT) which occurs, 
for example, in three level atoms 
driven by two laser fields \cite{Harris97a,Scullybook} and with 
phenomena where intrinsic transparency occurs via decay interference \cite{Harris89a,Zhou97a,Paspalakis99a}. 

The atomic system under consideration is shown in figure 1a. 
It consists of three atomic levels, labelled
$|n \rangle$, $(n=0,1,2)$, with $\omega_{0} < \omega_{2} < \omega_{1}$, where $\omega_{n}$ denotes the energy of each atomic state.  
The atom is assumed to be initially in state $|0\rangle$.
The transition $|1\rangle \leftrightarrow |2\rangle$ is taken 
to be near-resonant with a photonic band edge, while
the transition $|0\rangle \leftrightarrow |1\rangle$ is assumed to be far 
away from the gap and can therefore be treated as occurring in free space. 
The Hamiltonian which describes the 
dynamics of this system, in the interaction picture and the rotating wave approximation, is given by
$(\hbar = 1)$,
\begin{eqnarray} 
H &=& \bigg[ \Omega e^{i \delta t} |0\rangle \langle 1| + \sum_{{\bf k},\lambda}g_{{\bf k},\lambda}e^{-i(\omega_{\bf k}- \omega_{12})t} |1\rangle \langle2| \alpha_{{\bf k},\lambda}  \nonumber \\
&+& \mbox{H.c.} \bigg]
-i \frac{\gamma}{2} |1\rangle \langle1| \, . \label{Ham}
\end{eqnarray}
Here, $\Omega = - $\boldmath${\mu}$\unboldmath $_{01}\cdot$\boldmath${\epsilon}$\unboldmath $E$ is the Rabi frequency and $\delta = \omega
- \omega_{10}$, with $\omega_{nm} = \omega_{n} - \omega_{m}$, is the laser detuning from resonance of the $|0\rangle \leftrightarrow |1\rangle$ transition. In addition, $g_{{\bf k},\lambda} = - i \sqrt{2 \pi \omega_{\bf k}/V}$ \boldmath$\epsilon$\unboldmath $_{{\bf k},\lambda}\cdot$\boldmath$\mu$\unboldmath$_{12}$ 
denotes the coupling of the atom with the
modified vacuum modes. Both the Rabi frequency and the atom-vacuum coupling
strength are taken to be real. The dipole matrix element
of the $|n\rangle \leftrightarrow |m\rangle$ transition is
denoted by \boldmath$\mu$\unboldmath$_{nm}$. Also, \boldmath$\epsilon \, $\unboldmath and $E$ are 
respectively the polarization unit vector and electric field
amplitude of the laser field, while
\boldmath$\epsilon$\unboldmath$_{{\bf k},\lambda}$ is the polarization
unit vector, $\alpha_{{\bf k},\lambda}$ is the photon
annihilation operator, $\omega_{\bf k}$ is the
angular frequency of the $\{{\bf k},\lambda\}$ mode of the quantized vacuum 
field and $V$ is the quantization volume. Finally, $\gamma$ denotes the background
decay rate from state $|1\rangle$ to all other states of the atom. It is
assumed that these states are situated far from the
gap so that such background decay can be treated as a Markovian processes. 
The radiative shift associated with this decay has been omitted.
We note that, as long as the laser field is 
sufficiently weak, $\gamma$ can also account for the 
radiative decay of state $|1\rangle$ to state $|0\rangle$.

We proceed by expanding the wavefunction of the system, 
at a specific time $t$, in terms of the
`bare' state vectors such that
\begin{eqnarray}
 |\psi(t)\rangle = a_{0}(t)|0,\{0\}\rangle + a_{1}(t) e^{-i \delta t} |1,\{0\} \rangle \nonumber \\
 + \sum_{{\bf k},\lambda}a_{{\bf k},\lambda}(t)|2,\{{\bf k},\lambda\}\rangle \, . \label{wav}
\end{eqnarray}
Substituting Eqs.\ (\ref{Ham}) and (\ref{wav}) into the time-dependent Schr\"{o}dinger equation
and eliminating the vacuum amplitude $a_{{\bf k},\lambda}(t )$, we obtain
\begin{eqnarray}
i\dot{a}_{0}(t) &=& \Omega a_{1}(t) \label{a0} \, \\
i\dot{a}_{1}(t) &=& \Omega a_{0}(t) - \left(\delta + i \frac{\gamma}{2}\right)a_{1}(t) \nonumber \\
&-& i \int^{t}_{0}dt^{\prime}K(t-t^{\prime})a_{1}(t^{\prime}) \, , \label{a1} 
\end{eqnarray}
with the kernel
\begin{equation}
K(t-t^{\prime}) = \sum_{{\bf k},\lambda}  g^{2}_{{\bf k},\lambda} e^{-i(\omega_{\bf k}- \omega_{12}-\delta) (t-t^{\prime})} \, . \label{kerneltot}
\end{equation}
For the case of a Markovian reservoir
$K(t-t^{\prime}) = (\gamma_{1}/2)\delta(t - t^{\prime})$ with 
$\gamma_{1}$ being the decay rate to the state $|2\rangle$. 
However, for the case
of an {\it isotropic} model of the PBG which we consider here,
an effective mass 
dispersion relation \cite{Quang94a,Bay97a,Vats98a}
$\omega_{\bf k} = 
\omega_{g} + A(|{\bf k}| - |{\bf k}_{0}|)^2$, with $A \approx \omega_{g}/ |{\bf k}_{0}|^2$ is used, so that one obtains for the kernel
\begin{equation}
K(t-t^{\prime}) = \frac{\beta^{3/2} e^{-i[\pi/4 + (\delta_{g} - \delta)(t - t^{\prime})]}}{\sqrt{\pi(t - t^{\prime})}}, \quad t > t^{\prime}
\, , \label{kernel}
\end{equation}
with $\beta^{3/2} = 2\omega^{7/2}_{12}|$\boldmath$\mu$\unboldmath$_{12}|^{2}/(3c^{3})$
and $\delta_{g} = \omega_{g} - \omega_{12}$.
The isotropic dispersion relation leads to an
inverse square root density of modes for the modified
reservoir $\rho(\omega) \sim \Theta(\omega - \omega_{g})/\sqrt{\omega - \omega_{g}}$,
with $\Theta$ being the Heaviside step function (see figure 1b). 
We note that 
a similar density of modes is also found in waveguides \cite{Kleppner81a}
and in microcavities \cite{Lewenstein88a}, 
so that our results apply to these cases as well.

The aim here is to investigate the absorption and dispersion
properties of our system for a {\it weak} probe laser field. The equation of motion
for the electric field amplitude $E(z,t)$ is given by \cite{Harris92a},
\begin{eqnarray}
\left( \frac{\partial }{\partial z} + \frac{1}{v_{g}}\frac{\partial }{\partial t} \right) E(z,t) = - i \frac{\omega}{2c} \chi(\delta) E(z,t) \, ,
\end{eqnarray}
where $\chi(\delta)$ is the steady state
linear susceptibility of the medium and 
$v_{g} = c/[1 + (\omega/2)(\partial \mbox{Re}(\chi)/\partial \omega)]$ is
the group velocity of the laser pulse with the derivative of the real
part of the susceptibility being evaluated at the carrier frequency.

Since the transition $|0\rangle \leftrightarrow |1\rangle$ is treated as 
occurring in free space, the steady state linear susceptibility is given by
\cite{Scullybook} 
\begin{equation}
\chi(\delta) = - \frac{4\pi {\cal N} |\mbox{\boldmath${\mu}$\unboldmath $_{01}$}|^{2}}{\Omega}a_{0}(t \rightarrow \infty)a^{*}_{1}(t \rightarrow \infty) \, ,
\end{equation}
with ${\cal N}$ being the atomic density.
The solution of Eqs. (\ref{a0}) and (\ref{a1}) 
is obtained by means of perturbation theory \cite{Harris97a,Scullybook,Harris89a}. 
We assume that the laser-atom interaction is
very weak $(\Omega \ll \beta, \gamma)$ so that $a_{0}(t) \approx
1$ for all times. With the use of the Laplace transform we obtain from Eq.\ (\ref{a1})
\begin{equation}
A_{1}(s) = \frac{\Omega}{s\left[\delta +i\gamma/2 + i\tilde{K}(s) + is \right]} \, , \label{A1}
\end{equation}
where $A_{1}(s) = {\cal L}[a_{1}(t)]$, 
$\tilde{K}(s) = {\cal L}[K(t)]$ and $s$ 
is the Laplace variable. 
The inversion $a_{1}(t) = {\cal L}^{-1}[A_{1}(s)]$ 
is cumbersome and will not be presented
here. If $\gamma \neq 0$ then the terms inside the brackets of Eq. (\ref{A1}) have only complex, not purely imaginary, roots. 
Therefore we can easily obtain, using the final 
value theorem, the long time behaviour of the probability amplitude,
\begin{eqnarray}
a_{1}(t \rightarrow \infty) &=& \lim_{s \rightarrow 0}[sA_{1}(s)] \nonumber \\
&=& \frac{\Omega}{\delta + i \gamma/2 + i \tilde{K}(0)} \, . 
\end{eqnarray}
For the isotropic PBG model, using Eq.\ (\ref{kernel}), 
\begin{equation}
\tilde{K}(s) = \frac{\beta^{3/2}e^{-i\pi/4}}{\sqrt{s+i(\delta_{g}-\delta)}}
\, , 
\end{equation}
and the linear susceptibility reads 
\begin{eqnarray}
\chi(\delta) \sim \left\{ \begin{array}{rll} 
-\frac{\sqrt{\delta_{g}-\delta}}{(\delta - i\gamma/2) \sqrt{\delta_{g}-\delta} + \beta^{3/2}} \quad  \mbox{for} \quad  \delta \le \delta_{g} && \\
&& \\    \label{xpbg}
-\frac{\sqrt{\delta-\delta_{g}}}{(\delta - i\gamma/2) \sqrt{\delta-\delta_{g}} - i\beta^{3/2}} \quad  
\mbox{for} \quad \delta > \delta_{g} && \end{array} \right. . 
\end{eqnarray}
We see that if $\delta = \delta_{g}$, 
then $\chi(\delta_{g}) = 0$ and the system 
becomes transparent
to the laser field. In the case that the threshold frequency of the
band edge is equal to the  $|1\rangle \leftrightarrow |2\rangle$ 
transition frequency ($\delta_{g} = 0$), transparency 
occurs when the laser is
on resonance, i.e. at $\delta = 0$. This result is in contrast with
the case in which the transition $|1\rangle \leftrightarrow |2\rangle$ 
occurs in free space, where the well-known Lorentzian 
absorption profile \cite{Scullybook} is
obtained. In figure 2 we plot the linear 
absorption and dispersion spectrum of our system for
different values of the detuning 
of the atomic transition
$|1\rangle \leftrightarrow |2\rangle$ from the band edge threshold. 
Both the absorption and 
dispersion spectra are asymmetric and their shape depends critically
on this detuning. 

The group velocity of the pulse
can also exhibit interesting properties due to the steepness
of the dispersion curve. Unusually small group velocities in atomic vapors
have been predicted \cite{Harris92a}
and recently observed by several groups \cite{Schmidt96a,Hau99a,Kash99a}
in the phenomenon of EIT.
In our system the derivative of the real part of the susceptibility
diverges as the transparency condition $\delta = \delta_{g}$ is approached 
from below, leading to extremely slow group velocities, $v_{g} \rightarrow 0$.

We note that the transparency condition $\delta = \delta_{g}$ is 
similar to the two-photon resonance condition that leads to EIT
in ${\bf \Lambda}$-type atoms \cite{Harris97a,Scullybook}. However,
EIT occurs through the
application of two laser fields: one strong, coupling laser field
and one weak, probe laser field. Here
transparency is intrinsic to the system
as it occurs due to the presence
of a square root singularity at the density of modes threshold. 

Up to now, we have discussed the case
of an isotropic model for the PBG. We can also investigate
an {\it anisotropic} model of the PBG, where the dispersion
relation is given by \cite{John95a,Quang94a,Vats98a}
$\omega_{\bf k} = \omega_{g} + A({\bf k} - {\bf k}_{0})^2$.
In this case the associated density of modes 
near the edge of the PBG has a square root
threshold behaviour, $\rho(\omega) \sim \Theta(\omega - \omega_{g})\sqrt{\omega - \omega_{g}}$. The kernel of Eq. (\ref{kerneltot})
for the anisotropic PBG model is given by
\cite{John95a,Quang94a,Vats98a},
\begin{eqnarray}
K_{a}(t - t^{\prime}) \approx \frac{\beta_{a}^{1/2} 
e^{i[\pi/4 - (\delta_{g} - \delta)(t - t^{\prime})]}}{\sqrt{\pi}(t - t^{\prime})^{3/2}} \, ,
\nonumber \\
 \mbox{for} \quad \omega_{g}(t - t^{\prime}) \gg 1 
\, , \label{kernel2}
\end{eqnarray}
with $\beta^{1/2}_{a} = \omega^{2}_{12}|$\boldmath$\mu$\unboldmath$_{12}|^{2}/(2\omega_{g}A^{3/2})$.  Therefore $\tilde{K}_{a}(s) \sim \sqrt{s + i(\delta_{g}-\delta)}$
so the linear susceptibility does not go to zero for
any value of the probe detuning
and transparency does not occur in the anisotropic PBG model.

In summary, we have shown that a ${\bf \Lambda}$-type atom in which
one transition spontaneously
decays near the edge of an isotropic PBG 
can become transparent to a weak laser
field. Studies of quantum optical processes occurring in 
atoms embedded in PBG materials have, to date,
concentrated on the spontaneous emission dynamics \cite{John95a,Yablonovitch87a,Quang94a,Kurizki94a,Bay97a,Quang97a,Zhu97a,Bay97b,Vats98a,Kofman96a}. 
Our results suggest that the absorption and dispersion dynamics
of such atoms could reveal many surprising effects,
in particular in connection with other quantum coherence
and interference phenomena such as, for example, lasing without inversion and
nonlinear processes involving transparency \cite{Harris97a,Scullybook}.

This work has been supported in part
by the United Kingdom Engineering and Physical 
Sciences Research Council (EPSRC) and the European Commission
Cavity QED TMR Network ERBFMRXCT96066.

\pagebreak

\begin{figure}
\centerline{\psfig{figure=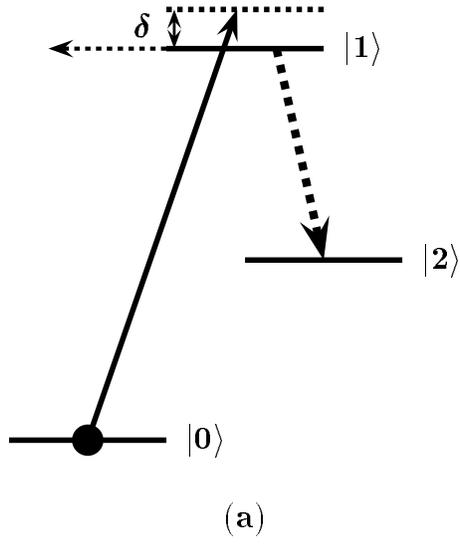,width=4.5cm}}

\vspace*{1.cm}

\centerline{\psfig{figure=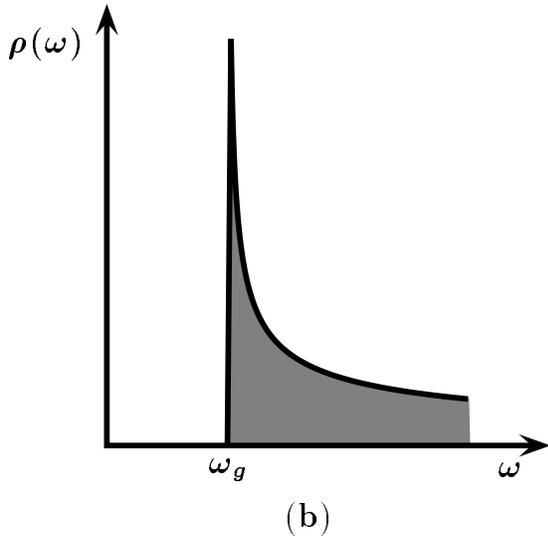,width=7.cm}}
\centerline{
\caption
{\narrowtext  
Figure (a) displays a three level, ${\bf \Lambda}$-type 
atomic system. The solid 
line denotes the probe laser coupling, the 
thick dashed line
denotes the coupling to the modified reservoir (PBG) and
finally the thin dashed line denotes the background
decay. Figure (b) shows the density of modes for the case 
of the isotropic PBG model.}}
\label{fig1}
\end{figure}

\pagebreak

\begin{figure}
\centerline{\psfig{figure=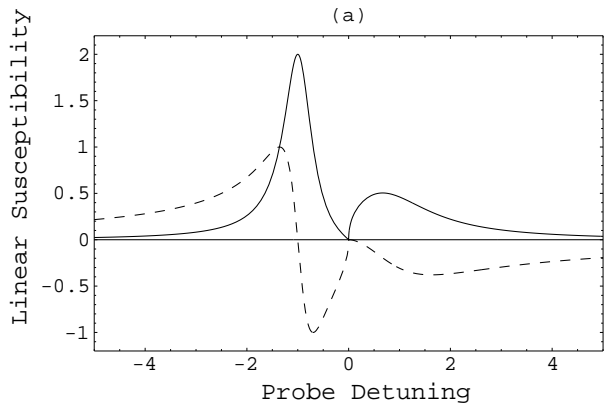,height=5.5cm}}

\centerline{\psfig{figure=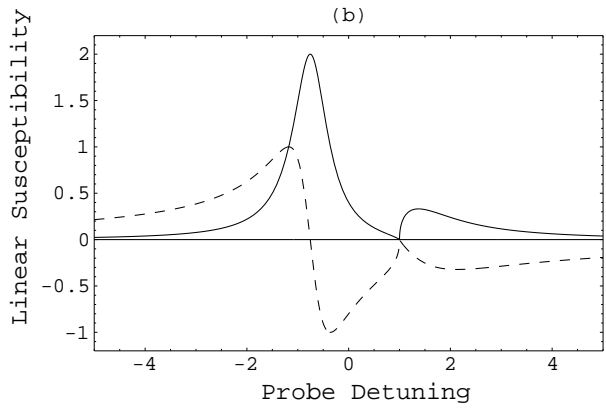,height=5.5cm}}

\centerline{\psfig{figure=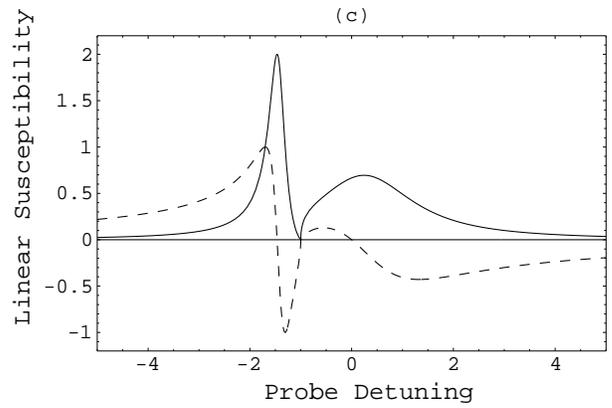,height=5.5cm}}
\centerline{
\caption
{\narrowtext The absorption and dispersion spectra (in arbitrary units) 
of our system
for parameters $\gamma = 1$, and (a) $\delta_{g} = 0$; (b)
$\delta_{g} = 1$; (c) $\delta_{g} = -1$. 
All parameters are in units of $\beta$. 
The solid curve is the absorption
profile ($-$Im$[\chi(\delta)]$) while the dashed 
curve the dispersion
profile (Re$[\chi(\delta)]$).}}
\label{fig2}
\end{figure}

\end{multicols}

\end{document}